\documentclass[aps,superscriptaddress,showpacs,nofootinbib,twocolumn]{revtex4}
\usepackage{bm}
\usepackage{xcolor}
\usepackage{graphicx,color}
\usepackage{amsmath}
\usepackage{amssymb}
\usepackage{soul}
\usepackage{booktabs}

\newcommand{\be}{\begin{equation}}
\newcommand{\ee}{\end{equation}}
\newcommand{\bea}{\begin{eqnarray}}
\newcommand{\eea}{\end{eqnarray}}

\begin{document}

%--------------------------------------------------------------------------------
\title{Homogeneous isotropic turbulence in four spatial dimensions}
%--------------------------------------------------------------------------------

\author{Arjun Berera}
%\email{ab@ph.ed.ac.uk}
\affiliation{School of Physics,
University of Edinburgh, Edinburgh EH9 3JZ, United Kingdom}
\author{Richard D. J. G. Ho \footnote{Present address: Marian Smoluchowski Institute of Physics, 
Jagiellonian University, {\L}ojasiewicza 11, 30-348, Krak{\'o}w, Poland}}
%\email{richard.ho@ed.ac.uk}
\affiliation{School of Physics,
University of Edinburgh, Edinburgh EH9 3JZ, United Kingdom}
\author{Daniel Clark}
%\email{D.Clark-9@sms.ed.ac.uk}
\affiliation{School of Physics,
University of Edinburgh, Edinburgh EH9 3JZ, United Kingdom}

%\author{}
%\email{}
%\affiliation{}

\begin{abstract}
Direct Numerical Simulation is performed of the forced Navier-Stokes equation
in four spatial dimensions. Well equilibrated, long time runs at
sufficient resolution were obtained to reliably measure spectral quantities, the velocity derivative skewness and the dimensionless dissipation rate. Comparisons to corresponding two and three dimensional results are made.
Energy fluctuations are measured and show a clear reduction moving from three to four dimensions. The dynamics appear to show simplifications
in four dimensions with a picture of increased forward energy transfer resulting in an extended inertial range with smaller Kolmogorov scale.
This enhanced forwards transfer is linked to our finding of increased
dissipative anomaly and velocity derivative skewness. 
\end{abstract}


\date{\today}

\maketitle

\section{Introduction}
\label{intro}

Turbulence is considered the oldest unsolved problem of theoretical physics. 
Moreover, the difficulty of the problem is such that 
it is still unknown if solutions to the underlying three dimensional 
Navier-Stokes equations (NSE) can exhibit singularities in finite time. 
In recent years, major 
advances have been made in understanding the turbulent behavior of the 
NSE using direct numerical simulation (DNS). In some ways this success 
has diverted efforts in physics from understanding the underlying 
structure of this equation in which a solution to the problem of turbulence 
may lie.  

In this paper, we shed new light on the properties of the NSE by utilizing DNS to study this 
equation in four spatial dimensions.  
A common tool of theoretical physics is to examine systems under different conditions, 
in our case the spatial dimension, in order to obtain new insights. 
We report the first results of fully developed statistically
stationary turbulence in 4D, building on the work of Gotoh {\it et al.}\cite{gwsns07} who performed
simulations of free decay. Through a comprehensive dataset that spans a significant 
region of parameter space we aim to provide phenomenological insights into
the connection between turbulence dynamics and dimensionality
in the wider context of complex systems. 
Fluctuation and dissipation are quantities typically studied in understanding
complex systems of many degrees of freedom, and in this paper
we will make one new measurement of each in 4D, to add to our general knowledge about turbulence.

Before explaining the necessary technical details and the numerical results, we provide a 
summary of existing works on the subject of dimensionality 
in turbulence, with the aim of raising interest in the subject from the wider 
theoretical physics community. Investigations have been conducted for the mathematical structure of the NSE in higher spatial dimensions \cite{scheffer, dong1, dong2, dong3, guo}. 
In the study of fully developed turbulence, 
the role of the spatial dimension has been an area of sustained focus. 
This has been particularly true ever since the development of 
renormalization group methods and their successful application 
to critical phenomena by Wilson and Fisher \cite{wf72,wilson83,fisher98}. 
These problems, as well as quantum field theory, 
share a number of common features. 
Of central importance is the presence of a large number of degrees of 
freedom across a range of length scales, which are often strongly interacting. 
As such, fluctuations, which may be present over many of these length scales, 
play a crucial role in these systems. 
In the case of critical phenomena, the connection between 
dimensionality and the suppression of fluctuations was understood 
by the Wilson-Fisher fixed point \cite{wf72} at four spatial dimensions, 
though there was also earlier work, such as by Ginzburg \cite{ginzburg60}, 
pointing to the relevance of $D=4$.

The success of this work in the early 1970s led to the application of the 
same renormalization group ideas to many other systems in physics. For 
the case of turbulence, this dates back to the work by 
Forster, Nelson, and Stephens \cite{fns76,fns77} and 
DeDominicis and Martins \cite{dm79}. These initial works motivated 
renormalization-group techniques as a means to treat the multi-scale 
physics of turbulence.  Embedded in this approach, the spatial dimension 
parameter is a prevalent feature and many subsequent studies have examined 
how the fixed point properties depend on it \cite{fournier, yakhot, teodorovich, eyink1, zhou, berera1}.

The connection between turbulence and quantum field theory
extends beyond just the development of the renormalization group.
In early work by Kraichnan \cite{kraichnan59},
Wyld \cite{wyld61}, and Edwards \cite{edwards64},
quantum field theory methods were utilized
to develop a perturbation theory for the Navier-Stokes equation. 
Subsequently, notable works by Martin, Siggia, and Rose \cite{msr73} using a Hamiltonian
approach, and by Jensen \cite{jensen81}
using a functional integral approach, and 
many others \cite{kawasaki, mcomb1, deker, phythian, janssen, andersen, berera2, frederiksen}, continued developing
perturbation expansions of the NSE, all borrowing ideas from quantum
field theory.

Associations between turbulence and
gauge theories have also been made.
Quantum Chromodynamics (QCD) is one example.
The confinement problem of QCD
has similarities to turbulence, due to both having many
degrees of freedom involving multi-scale physics, and
in addition both are strong coupling problems.
One of the first lattice gauge theory simulations by Creutz
examined the dependence 
of QCD on dimensionality \cite{Creutz:1979dw}, with
qualitative differences found in
confinement behavior in four versus three spatial dimensions.
The QCD connection to turbulence was greatly enhanced 
by the works
of Migdal \cite{migdal, migdal2, migdal3},
in developing an analog
for the Navier-Stokes equation
to the Wilson loop of gauge theories \cite{Wilson:1974sk},
and Polyakov \cite{polyakov1, polyakov2, polyakov3, polyakov4},
in using conformal field
theory (CFT) methods to study two-dimensional turbulence,
with connections also made between turbulence and the ADS/CFT
correspondence \cite{maldacena,fluidgrav,adscftturb}. 
From 
another direction, the scaling exhibited by the asymptotically
free ultraviolet behavior of QCD has been noted to
have similarities to scaling in turbulence \cite{eg1994}.
Also, the Galilean invariance of the Navier Stokes equation has
been interpreted as a gauge invariance \cite{abdh1, abdh2}.

Motivated from these various directions, there have been
many theoretical studies examining the dependence on spatial dimensionality of
turbulence.  Some have explicitly developed analogies
between turbulence and critical phenomena, and through that
the possibility of a critical dimension for turbulence 
\cite{eg1994, bramwell, aji, nelkin1, frisch1, yakhot2, lvov, giuliani, frisch3, n74, nb78, fsn78, ff78, lk97}, above which the Kolmogorov
theory (K41) \cite{k41} may become exact.
Following this line of reasoning,
studying turbulence between two and three spatial dimensions reveals a change in
energy cascade directions \cite{ff78,nelkin1,frisch1,yakhot2,lvov,giuliani,frisch3}.
This has been associated with a lower critical dimension existing at a non-integer intermediate
dimension close to $D=2$.
Numerical \cite{celani, benavides, alexakis1} and experimental \cite{xia2011} results show that cascade directions can indeed change as a function 
of different control parameters, one of which being the aspect ratio of the domain . Additionally, the behavior of passive scalars in higher dimensional turbulence has also been investigated \cite{kraichnan1974scalar}, and here it was found that for a certain prescribed velocity field intermittency vanished in the $d\rightarrow \infty$ limit.
Furthermore, studies above three dimensions
\cite{n74,nb78,fsn78,ff78,khesin,kraichnan1994a,ffr} 
have made various claims
as to the extent turbulent behavior changes at higher dimensions.
Kraichnan \cite{k85} and Meneveau and Nelkin \cite{mn89} predict a change in
inertial range behavior at spatial dimension $D=4$. Further to this,
Liao \cite{liao90, liao91} and Nelkin \cite{nelkin03} argue, through close analogies to critical phenomena, for
an upper critical spatial dimension of six and four respectively for turbulence.  
Similarities between the
NSE and the Kardar-Parisi-Zhang (KPZ) equation have
also been observed \cite{lk97} due to both having nonlinear strong
coupling regimes. In the latter,
it has been theoretically argued that four spatial dimensions is a type
of critical dimension, whether a corresponding result exists for the NSE is 
left as an interesting question.
More recently a study \cite{ffo10} of fluid velocity
correlation functions in varying dimensions highlighted competing
effects on the statistics. This work gave some analytic relations 
but left essential open questions requiring numerical study.

From this short review, it is clear that understanding
the dependence of turbulent behavior in the NSE on 
spatial dimension has been an area of sustained interest for at least the past half century. Theoretical considerations and speculations are abundant, with many analogies made to critical 
phenomena and quantum field theory, where it is already an established fact that spatial 
dimensionality plays a significant role in governing behavior.  
These theoretical treatments 
provide strong motivation to numerically study the properties of turbulence in spatial 
dimensions higher than three. This is a very computationally expensive challenge, though computing 
power has reached a stage where meaningful studies can now be performed.

Previous investigations into higher dimensional turbulence via DNS have been
insightfully motivated but limited in scope \cite{Suzuki2005,gwsns07,ysing12,nikitin11}.
All four of these studies were for freely decaying turbulence, with
short run times and relatively coarse grid spacing, a restriction imposed by the 
available computing power at that time.
The maximum grid resolution in any direction was 256
%which is known to be inadequate in 3D to see 
collocation points, which in 3D is not sufficient to result in 
a scaling region for free decay, 
with at least 512 being a safe minimum.
For example, the scaling reported in \cite{Suzuki2005,gwsns07,ysing12}
was based on normalization of the energy spectra, 
which has its ambiguities in capturing the scaling regime.  
Moreover the short simulation times in all
four papers \cite{gwsns07,ysing12,nikitin11} increases the risk of
being influenced by the effect of the initial conditions on the
statistics.
Nevertheless, these papers presented the first measurements of important observables in 
turbulence such as energy spectra, the skewness of velocity-field gradients and energy 
decay rates in 4D turbulence to the best possible accuracy permitted by computing power at that time.
All these groundbreaking simulations were done some time ago.
What is needed and possible now are larger, stationary state, simulations.
Although this is computationally intensive, it is necessary if
the data are to be reliable and not rely on external assumptions.
In this paper we are able to
go to large enough box size and evolution time in forced simulations
to report the first fully-developed turbulence datasets in 4D.
Moreover the past 4D DNS studies placed considerable focus on intermittency
properties. Our paper is following a different physical
motivation, that of turbulence as an example of
a strong coupling problem.  It is in that context we
presented examples in this Introduction from critical
phenomenon and quantum field theory, with turbulence
yet being another example of a strongly coupled theory.

\section{Basic equations}
\label{basice}

In this study we look systematically at forced DNS in two, three, and four spatial dimensions. In such simulations, a steady state is reached which 
allows for a clear scaling regime to be identified,
with statistics taken for multiple large eddy turnover times
and performed on up to $512^4$ collocation points. Obtaining such a large dataset is a non-trivial task, but is necessary
to reach a level where  the spectral
quantities and correlations typically associated
with turbulence can be reliably measured. 
By doing so it is possible to make direct comparison to turbulence in two and three spatial
dimensions.

The Navier-Stokes equations
\begin{equation}\label{NSE} 
\begin{split}
\partial_t u_i + u_j\Omega_{ji} &= -\partial_i\left( P + \frac{u^2}{2}\right) + \nu\nabla^2 u_i + f_i \ , \\
\partial_i u_i &= 0 \ ,
\end{split}
\end{equation} 
are numerically integrated using a fully de-aliased
pseudo-spectral code in a periodic cube of length $2\pi$ \cite{YoffeThesis,EddyBurgh}.
Here, $\bm{u}(\bm{x},t)$ is the velocity field, $P(\bm{x},t)$ is the pressure field, $\nu$ is 
the kinematic viscosity, $\bm{f}(\bm{x},t)$ is an external force and $\Omega_{ij} = \partial_i 
u_j - \partial_j u_i$ is the vorticity 2-form.
The density was set to unity. Equation (\ref{NSE})
is equivalent to the standard form in all dimensions.

For fluid flows of any dimension, inviscid invariants exist depending on
whether the spatial dimension is odd or even, referred to as helicity-type and
enstrophy-type invariants respectively \cite{invar4d}. Thus, to ensure the
correctness of our four-dimensional NSE implementation, we measured the lowest
order invariant and found it was indeed conserved in the non-linear term.

The primary forcing used was a negative damping scheme which 
only forced the low wave numbers (large scales), 
$k_f = 2.5$, according to the rule
\begin{equation}
\boldsymbol{\hat{f}}(\boldsymbol{k},t) =\begin{cases}
 (\varepsilon/2E_{f})\boldsymbol{\hat{u}}(\boldsymbol{k},t) \quad &\text{if } 0 < k \leq k_f, \\
 0 \qquad &\text{else,}
\end{cases}
\end{equation} 
where $E_f$ is the energy in the forcing band $0 < k \leq k_f$
and $\bm{\hat{u}}(\bm{k},t)$ is the Fourier transform of field $\bm{u}$.
This well tested forcing function \cite{force,Kaneda2006,Linkmann2015}
allows the dissipation rate, $\varepsilon$, to be known a priori.
We set $\varepsilon$ to 0.1 for all runs. 
The simulations were well resolved, with $k_{\text{max}} \eta > 1$ for
all simulations, where $k_{\text{max}}$ is the largest wavenumber in the 
simulation and $\eta$ the Kolmogorov microscale.
Simulations were initialized randomly from a
Gaussian distribution with zero mean.

The pseudo-spectral technique allows statistics of the field to be calculated
from the energy spectra.
Due to the properties of homogeneity and isotropy, the calculations depend on
the spatial dimension of the field.
In D-dimensional homogeneous isotropic turbulence $u$, the rms velocity, is defined as
$u = \sqrt{2E/D}$, 
where $D$ is the spatial dimension and $E$ the energy.
The integral length scale, $L_{D}$, and Taylor microscale, $\lambda_{D}$, are calculated from simulations as
\begin{equation} \begin{split}
L_D &= \frac{\Gamma(\frac{D}{2}) \sqrt{\pi}}{\Gamma(\frac{D+1}{2}) u^2}\int_{0}^{\infty} \mathrm{d}k \, E(k) k^{-1}\ ,  \\ \lambda_D &= \sqrt{\frac{D(D+2) \nu u^2}{\varepsilon}} \;,
\end{split}
\end{equation}
where $E(k)$ is the energy spectrum. The Reynolds numbers quoted throughout the paper are then the integral scale Reynolds number $\mathrm{Re}_{L} = uL_{D}/\nu$ and the Taylor Reynolds number $\mathrm{Re}_\lambda$ = $u\lambda_D /\nu$. Due to their dependence on the spatial dimension it is important the correct form is used, particularly for determining the scaling properties of the velocity derivative skewness as well as for measuring the correct value for the dimensionless dissipation rate.
%Together, these dependencies on spatial dimension mean that identical spectra
%for 3D and 4D, with the same $\nu$, would result in a roughly 10\% 
%higher value of $\mathrm{Re}_{L}$ for the 4D data.

\begingroup
\begin{table}[]
\squeezetable
\begin{tabular}{lllllllll}
\hline
Re$_L$ & Re$_{\lambda}$ & $T_0$  & $\nu$ & $L$ & $\lambda$ & $U$ & $k_{\mathrm{max}}$ & $\eta$   \\ \hline
160   & 74   & 2.31 & 0.0008  & 0.54 & 0.080 & 0.24 & 169  & 0.0842 \\
225   & 94   & 1.80 & 0.0008  & 0.57 & 0.080 & 0.32 & 169  & 0.0824 \\
248   & 72   & 5.18 & 0.0002  & 0.51 & 0.039 & 0.10 & 340  & 0.0350 \\
276   & 108  & 1.56 & 0.0008  & 0.59 & 0.080 & 0.38 & 169  & 0.0692 \\
319   & 124  & 1.37 & 0.0008  & 0.59 & 0.080 & 0.43 & 169  & 0.0684 \\
358   & 131  & 1.29 & 0.0008  & 0.61 & 0.080 & 0.47 & 169  & 0.0604 \\
626   & 171  & 1.68 & 0.0003  & 0.56 & 0.049 & 0.33 & 340  & 0.0456 \\
633   & 155  & 2.34 & 0.0002  & 0.54 & 0.038 & 0.23 & 340  & 0.0245 \\
676   & 167  & 2.16 & 0.0002  & 0.54 & 0.040 & 0.25 & 340  & 0.0420 \\
696   & 139  & 3.78 & 0.0001  & 0.51 & 0.027 & 0.14 & 340  & 0.0206 \\
698   & 186  & 1.54 & 0.0003  & 0.57 & 0.049 & 0.37 & 340  & 0.0404 \\
823   & 193  & 1.85 & 0.0002  & 0.55 & 0.040 & 0.30 & 340  & 0.0385 \\
945   & 217  & 1.62 & 0.0002  & 0.55 & 0.040 & 0.34 & 340  & 0.0397 \\
966   & 188  & 2.77 & 0.0001  & 0.52 & 0.028 & 0.19 & 340  & 0.0378 \\
984   & 196  & 1.78 & 0.0002  & 0.59 & 0.040 & 0.33 & 340  & 0.0271 \\
1054  & 231  & 1.53 & 0.0002  & 0.57 & 0.040 & 0.37 & 340  & 0.0355 \\
1157  & 242  & 1.45 & 0.0002  & 0.58 & 0.040 & 0.40 & 340  & 0.0320 \\
1180  & 223  & 2.32 & 0.0001  & 0.52 & 0.028 & 0.23 & 340  & 0.0297 \\
1318  & 225  & 2.29 & 0.0001  & 0.55 & 0.028 & 0.24 & 340  & 0.0192 \\
1360  & 246  & 2.09 & 0.0001  & 0.53 & 0.028 & 0.26 & 340  & 0.0284 \\
1488  & 259  & 1.98 & 0.0001  & 0.54 & 0.028 & 0.27 & 340  & 0.0275 \\
1659  & 277  & 1.82 & 0.0001  & 0.55 & 0.028 & 0.30 & 340  & 0.0286 \\
1696  & 268  & 2.13 & 0.00008 & 0.54 & 0.025 & 0.25 & 681  & 0.0271 \\
1915  & 304  & 1.67 & 0.0001  & 0.57 & 0.028 & 0.34 & 340  & 0.0257 \\
1954  & 276  & 2.70 & 0.00005 & 0.51 & 0.020 & 0.19 & 681  & 0.0262 \\
1985  & 297  & 1.69 & 0.0001  & 0.58 & 0.027 & 0.34 & 340  & 0.0154 \\
2026  & 336  & 1.51 & 0.0001  & 0.55 & 0.028 & 0.37 & 340  & 0.0265 \\
2241  & 350  & 1.44 & 0.0001  & 0.57 & 0.028 & 0.39 & 340  & 0.0257 \\
3274  & 352  & 2.73 & 0.00003 & 0.52 & 0.015 & 0.19 & 681  & 0.0176 \\
3900  & 378  & 2.78 & 0.000025 & 0.52 & 0.014 & 0.19 & 681  & 0.0168 \\
4925  & 435  & 2.69 & 0.00002 & 0.52 & 0.013 & 0.19 & 681  & 0.0151 \\
9831  & 610  & 2.73 & 0.00001 & 0.52 & 0.009 & 0.19 & 681  & 0.0106 \\
19485 & 878  & 2.69 & 0.000005 & 0.51 & 0.006 & 0.19 & 1364 & 0.0069 \\ \hline
\end{tabular}
\caption{Simulation parameters for 2D runs. Due to the inverse energy cascade a hypoviscous term, proportional to $\nabla^{-2}$, was utilized to prevent condensation at the largest scales. For all runs the sum of large and small scale energy dissipation was 0.1. Here $\eta = \left(\nu^3/\varepsilon_{\omega}\right)^{1/6}$, where $\varepsilon_{\omega}$ is the enstrophy dissipation rate.}
\label{tab:2d}
\end{table}
\endgroup

\begingroup
\begin{table}[]
\squeezetable
\begin{tabular}{lllllllll}
\hline
Re$_L$ & Re$_{\lambda}$ & $T_0$  & $\nu$ & $L$ & $\lambda$ & $U$ & $k_{\mathrm{max}}$ & $\eta$   \\ \hline
11   & 9    & 4.81 & 0.08    & 2.10 & 1.51 & 0.44 & 20   & 0.2675 \\
11   & 9    & 4.78 & 0.09    & 2.20 & 1.69 & 0.46 & 20   & 0.2922 \\
13   & 11   & 4.60 & 0.07    & 2.08 & 1.46 & 0.45 & 20   & 0.2420 \\
15   & 11   & 4.29 & 0.06    & 1.93 & 1.35 & 0.45 & 20   & 0.2156 \\
23   & 16   & 3.71 & 0.04    & 1.85 & 1.22 & 0.50 & 20   & 0.1591 \\
30   & 20   & 3.32 & 0.03    & 1.74 & 1.11 & 0.52 & 20   & 0.1282 \\
45   & 27   & 2.97 & 0.02    & 1.63 & 0.95 & 0.55 & 20   & 0.0946 \\
46   & 29   & 2.94 & 0.02    & 1.64 & 0.97 & 0.56 & 169  & 0.0946 \\
75   & 39   & 2.42 & 0.01    & 1.34 & 0.68 & 0.55 & 20   & 0.0562 \\
88   & 44   & 2.39 & 0.009   & 1.37 & 0.67 & 0.57 & 41   & 0.0520 \\
88   & 44   & 2.57 & 0.01    & 1.50 & 0.72 & 0.58 & 169  & 0.0562 \\
96   & 47   & 2.32 & 0.008   & 1.33 & 0.63 & 0.57 & 41   & 0.0476 \\
112  & 51   & 2.28 & 0.007   & 1.34 & 0.60 & 0.59 & 41   & 0.0430 \\
130  & 57   & 2.19 & 0.006   & 1.31 & 0.57 & 0.60 & 41   & 0.0383 \\
147  & 61   & 2.02 & 0.005   & 1.22 & 0.52 & 0.60 & 169  & 0.0334 \\
153  & 63   & 2.15 & 0.005   & 1.28 & 0.52 & 0.60 & 41   & 0.0334 \\
201  & 73   & 2.12 & 0.004   & 1.31 & 0.48 & 0.61 & 41   & 0.0283 \\
249  & 83   & 2.05 & 0.003   & 1.24 & 0.41 & 0.60 & 84   & 0.0228 \\
344  & 100  & 1.82 & 0.002   & 1.12 & 0.34 & 0.62 & 169  & 0.0168 \\
378  & 103  & 1.98 & 0.0019  & 1.19 & 0.32 & 0.60 & 84   & 0.0162 \\
393  & 107  & 2.05 & 0.002   & 1.27 & 0.34 & 0.62 & 84   & 0.0168 \\
395  & 108  & 1.94 & 0.0018  & 1.17 & 0.31 & 0.61 & 84   & 0.0155 \\
436  & 113  & 1.99 & 0.0017  & 1.22 & 0.31 & 0.61 & 84   & 0.0149 \\
484  & 120  & 2.02 & 0.0016  & 1.25 & 0.30 & 0.62 & 84   & 0.0142 \\
488  & 121  & 1.96 & 0.0015  & 1.20 & 0.29 & 0.61 & 84   & 0.0136 \\
536  & 125  & 1.98 & 0.0014  & 1.22 & 0.28 & 0.62 & 84   & 0.0129 \\
806  & 158  & 2.01 & 0.001   & 1.27 & 0.24 & 0.63 & 169  & 0.0100 \\
979  & 174  & 1.97 & 0.0008  & 1.24 & 0.22 & 0.63 & 169  & 0.0085 \\
1096 & 180  & 1.85 & 0.0006  & 1.10 & 0.18 & 0.60 & 169  & 0.0068 \\
1446 & 212  & 1.86 & 0.0005  & 1.16 & 0.17 & 0.62 & 340  & 0.0059 \\
2517 & 286  & 1.87 & 0.0003  & 1.19 & 0.13 & 0.64 & 340  & 0.0041 \\
6207 & 453  & 1.75 & 0.00011 & 1.09 & 0.08 & 0.63 & 681  & 0.0019 \\ \hline
\end{tabular}
\caption{Simulations parameters for 3D runs, $\varepsilon = 0.1$ for all cases.}
\label{tab:3d}
\end{table}
\endgroup

\begingroup
\begin{table}[]
\squeezetable
\begin{tabular}{lllllllll} 
\hline
Re$_L$ & Re$_{\lambda}$ & $T_0$  & $\nu$   & $L$    & $\lambda$    & $U$    & $k_{\mathrm{max}}$ & $\eta$   \\ \hline
27  & 15  & 4.99 & 0.03   & 2.01 & 1.12 & 0.40 & 20   & 0.130 \\
39  & 20  & 4.42 & 0.02   & 1.87 & 0.95 & 0.42 & 20   & 0.096 \\
52  & 24  & 4.12 & 0.015  & 1.79 & 0.84 & 0.44 & 20   & 0.077 \\
74  & 31  & 3.74 & 0.01   & 1.66 & 0.70 & 0.45 & 41   & 0.057 \\
99  & 38  & 3.70 & 0.008  & 1.71 & 0.65 & 0.46 & 41   & 0.048 \\
126 & 44  & 3.50 & 0.006  & 1.63 & 0.56 & 0.47 & 41   & 0.038 \\
141 & 46  & 3.36 & 0.005  & 1.54 & 0.51 & 0.46 & 41   & 0.034 \\
203 & 57  & 3.28 & 0.0035 & 1.53 & 0.43 & 0.47 & 84   & 0.026 \\
347 & 77  & 3.13 & 0.002  & 1.47 & 0.33 & 0.47 & 84   & 0.017 \\
838 & 124 & 2.94 & 0.0008 & 1.41 & 0.21 & 0.48 & 169  & 0.008 \\ \hline
\end{tabular}
\caption{Simulations parameters for 4D runs, $\varepsilon = 0.1$ for all cases.}
\label{tab:4d}
\end{table}
\endgroup

\section{Results}
\label{results}

In total, we carried out 
10 simulations in 4D for $27 \leqslant \mathrm{Re}_{L} \leqslant 838$ on $64^4-512^4$ grid points, 
27 runs in 2D with $160 \leqslant \mathrm{Re}_{L} \leqslant 19485$ on $512^2-4096^2$ grid points,
and used a 3D dataset \cite{BereraPRL} containing of 33 runs with 
$11 \leqslant \mathrm{Re}_{L} \leqslant 6207$ on $64^3-2048^3$ grid points. For more detailed information about the simulations performed see Tables \ref{tab:2d}, \ref{tab:3d} and \ref{tab:4d} for 2D, 3D and 4D respectively. These tables should be compared with table 1 in \cite{gwsns07}, although caution may be needed as it is not clear if the dimensional corrections to the Taylor length scale have been considered there. Nonetheless, it is clear from these tables that the 4D simulations presented here are at a higher resolution and Re$_{\lambda}$ than any work to date. Furthermore, as we make use of a large scale forcing term our results are for statistically stationary turbulence, as such, we can be confident that our measurements pertain to fully developed turbulence which is not true of the decaying runs performed in \cite{gwsns07}. In achieving fully developed turbulence our results allow the nature of 4D turbulence, and how it differs from the 3D case, to be understood and allows for theoretical ideas to be tested reliably.

\begin{figure}
\includegraphics[width=\columnwidth]{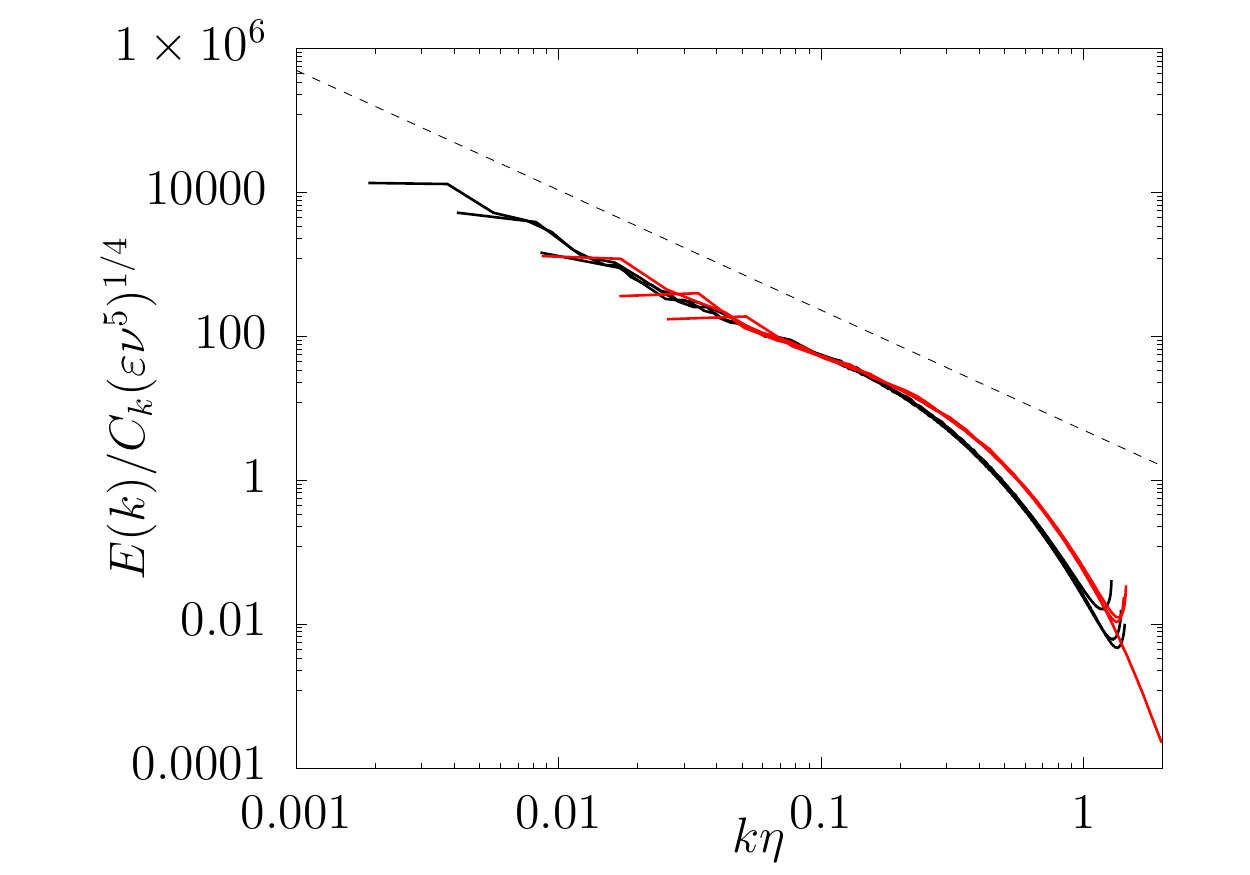}
\caption{\label{fig:plotEk} (Color online) Normalized energy spectrum, $E(k)$, against $k\eta$ at Re$_L$ = 6207, 2517 and 979 for 3D (black) and Re$_{L}$ = 838, 347 and 203 for 4D (red). Kolmogorov $k^{-5/3}$ predicted inertial range scaling (long-dashed black).}
\end{figure}

\begin{figure}
\includegraphics[width=\columnwidth]{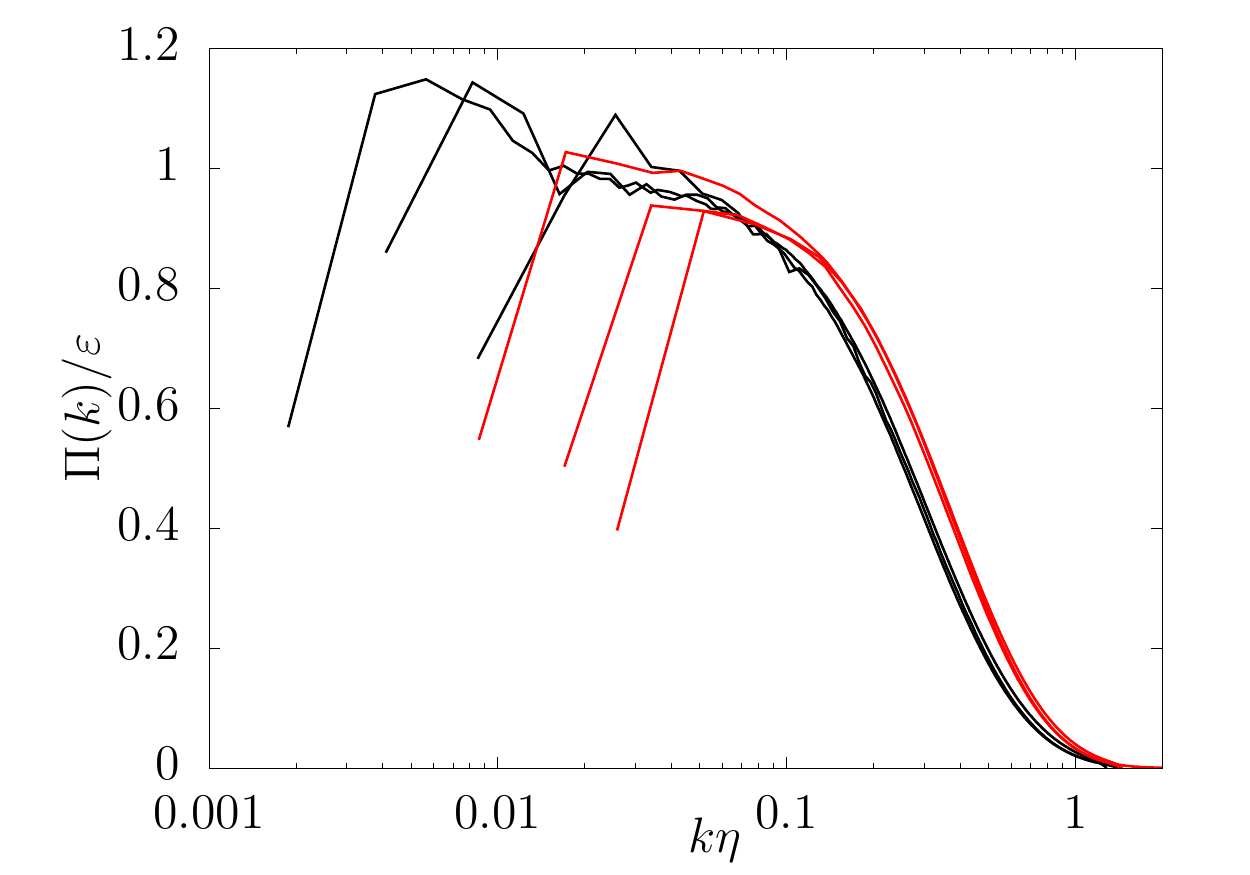}
\caption{\label{fig:plotFlux} (Color online) Normalized energy flux, $\Pi (k)$ against $k\eta$ at Re$_L$ = 6207, 2517 and 979 for 3D (black) and Re$_{L}$ = 838, 347 and 203 for 4D (red).}
\end{figure}

\begin{figure}
\includegraphics[width=\columnwidth]{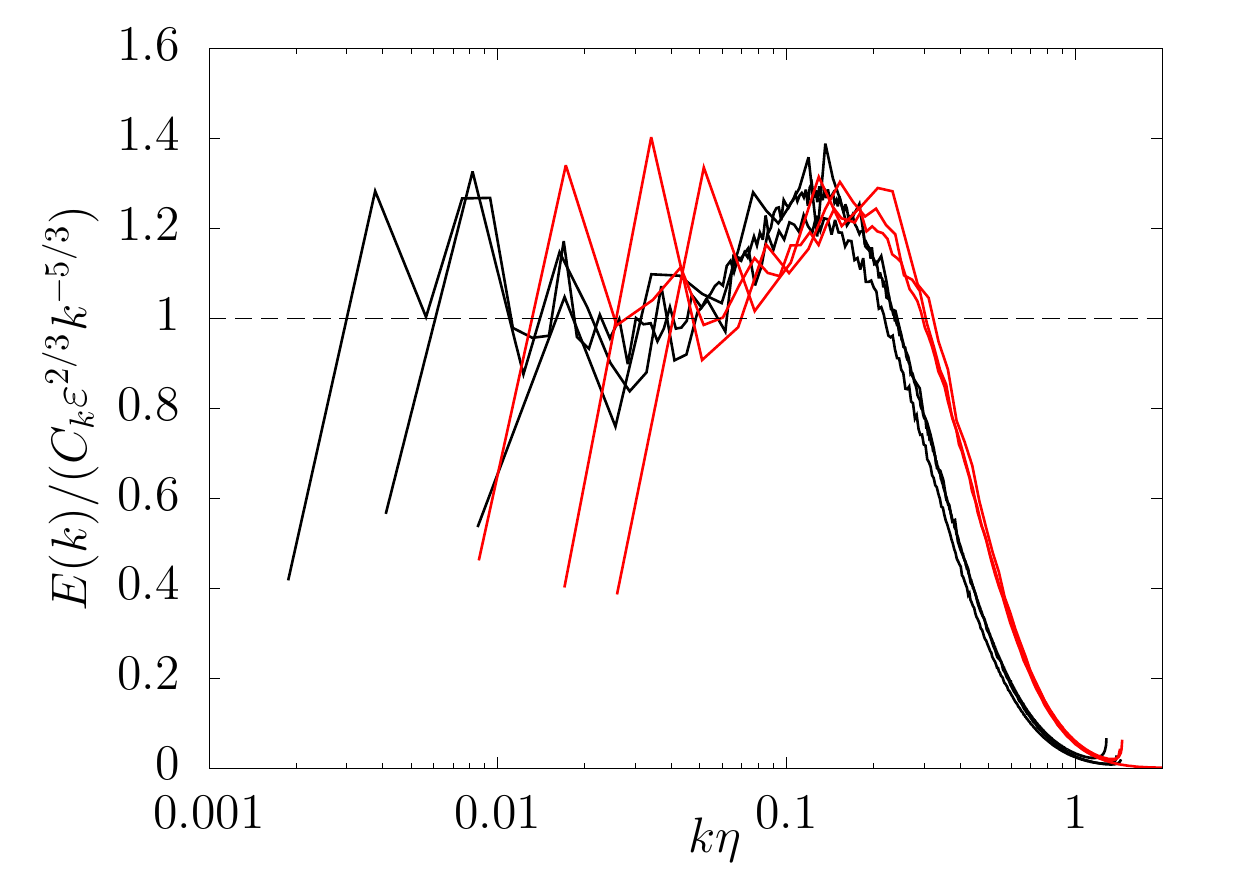}
\caption{\label{fig:plotEkcomplin} (Color online) Log-linear plot of compensated energy spectrum, $E(k)$, against $k\eta$ at Re$_L$ = 6207, 2517 and 979 for 3D (black) and Re$_{L}$ = 838, 347 and 203 for 4D (red). }
\end{figure}

%\begin{figure}
%\includegraphics[width=\columnwidth]{logCompensatedEnergyTex.pdf}
%\caption{\label{fig:plotEkcomplog} (Color online) Log-log plot of compensated energy %spectrum, $E(k)$, against $k\eta$ at Re$_L$ = 6207, 2517 and 979 for 3D (black) and Re%$_{L}$ = 838, 347 and 203 for 4D (red). }
%\end{figure}

From our simulations, we find that the energy and transfer spectra for 3D and 4D 
are very similar, and differ from those for 2D (we performed some decaying runs of 
4D turbulence and these showed no tendency towards inverse transfer, unlike that found in 2D).
%In Figure \ref{fig:plotEk} 
In Figures \ref{fig:plotEk} and \ref{fig:plotFlux}
we show $E(k)$ and $\Pi(k)$, the energy flux, respectively, 
for a set of 3D (with Re$_{L}$ from 980 to 6200) and 4D simulations (with Re$_{L}$ from 200 to 840),
which were taken from an ensemble of spectra over multiple large eddy turnover times $T_0 = L/u$,
at intervals longer than $T_0$.
%In this figure, the wave number $k$ is normalized by $\eta$, and $E(k)$ and $T(k)$
In Fig.~\ref{fig:plotEk}, the wave number $k$ is normalized by $\eta$, the energy spectra
are normalized with $\epsilon$ and $\nu$ such that they collapse on to the same
values in the dissipation range. However, as can be seen in the Figure,
the collapse of the spectra only apply if the spatial dimension is the same.
%Both have $E(k) \sim k^{-5/3}$ in the inertial range, which is in agreement with the 
%dimensional arguments given by Kolmogorov. 

Both 4D and 3D energy spectra are consistent with Kolmogorov $-5/3$ scaling, as can be 
seen in Fig.~\ref{fig:plotEk} and the compensated spectra shown in Fig.~\ref{fig:plotEkcomplin}. 
The scaling range in 4D is short, as 
can be expected when considering 3D data at comparable Reynolds numbers. However, we find the Kolmogorov constant, $C_k$, to be less in 4D than 3D
,consistent with \cite{gwsns07}, 
with 
%$C_{k}^{\mathrm{3D}} \approx 1.72$ and $C_{k}^{\mathrm{4D}} \approx 1.33$, 
$C_{k}^{\mathrm{3D}} \approx 1.7$ and $C_{k}^{\mathrm{4D}} \approx 1.3$, 
however, higher resolution simulations are needed to ascertain the true values \cite{ishihara2016high}.
One key difference, highlighted in both plots of the energy flux, Figure \ref{fig:plotFlux}, and compensated energy spectra, Figure \ref{fig:plotEkcomplin},
is the existence of a possibly extended scaling region in 4D as compared to 3D, as evidenced by the viscous sub-range beginning at a higher value of $k\eta$ in 4D.
This suggests that in this higher dimensional case, there is an increased forward transfer of 
energy, such that the effects of dissipation do not become dominant until at scales smaller 
than those in three dimensions. This increase of forward energy transfer with dimension is 
supported by theoretical predictions, where it is shown to be determined by the possible 
geometries of triad interactions as the dimension tends to infinity \cite{ff78}.
Our finding of a 
stronger forwards transfer is consistent with the larger decay exponent 
found in \cite{gwsns07}.
The extended scaling region is in agreement with \cite{k85},
which predicts 
$\eta$ being pushed to smaller values.

\begin{figure}
    \includegraphics[width=\columnwidth]{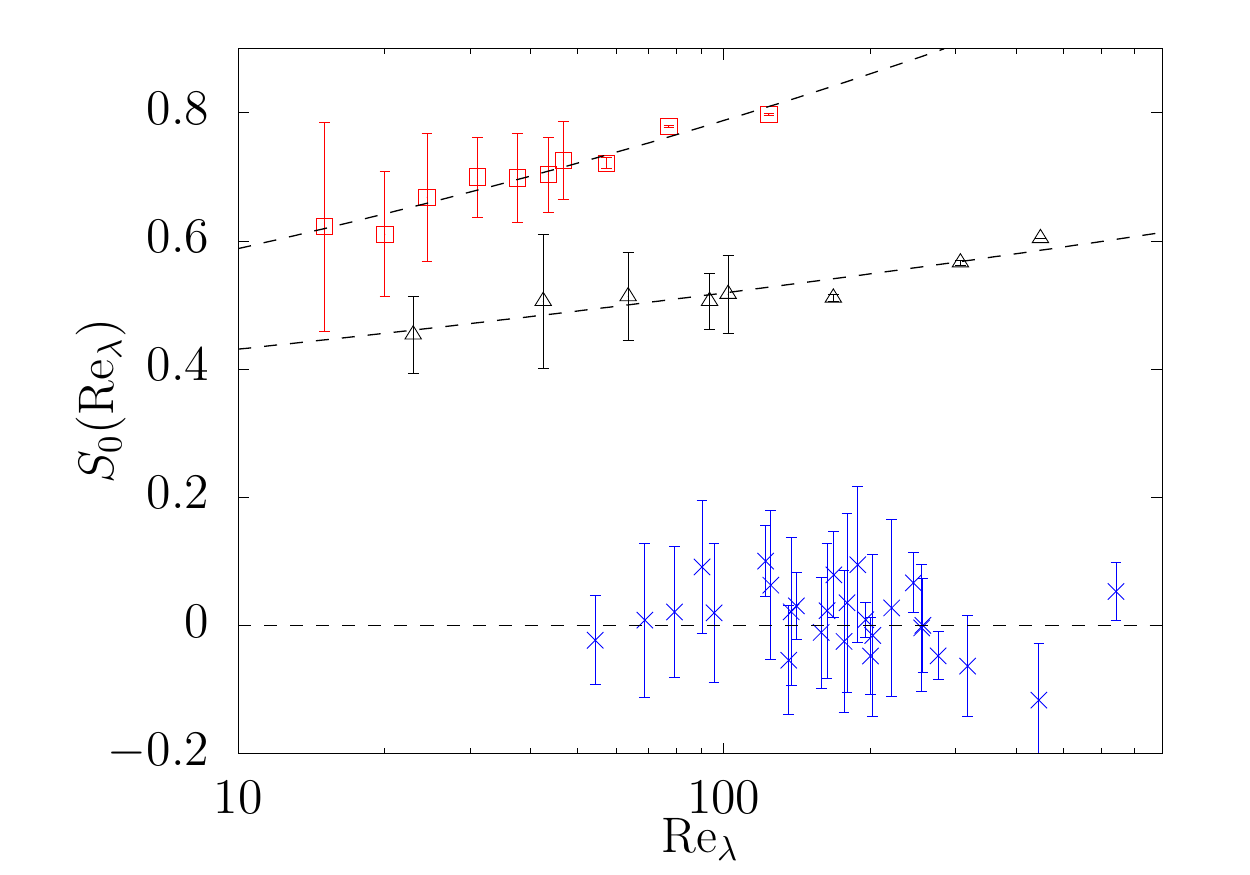}
    \caption{Velocity derivative skewness $S_0$ vs Re$_{\lambda}$ for two ($\times$), three ($\vartriangle$) and four dimensions ($\square$). The dashed line show least-squares fit to the power law $a$Re$_{\lambda}^{b}$ as predicted by Kolmogorov's refined similarity hypothesis. Though this hypothesis was originally formulated for 3D turbulence, given the forward energy cascade present in our 4D simulations it is likely a similar hypothesis could be presented in 4D. The error on the skewness was calculated for less time than other statistics for larger Re$_{\lambda}$}.
    \label{fig:skewness}
\end{figure}

To further investigate this enhanced forward transfer, we consider the 
von K\'arm\'an-Howarth equation. 
From this, it can be shown that the enstrophy equation in 3D
takes the form \cite{DavidsonBook}
\begin{equation}
\begin{split}
\partial_t Z(t) &= \frac{7}{3}\sqrt{\frac{2}{15}}S_0 Z(t)^{\frac{3}{2}} - 2 \nu P(t), \\  S_0 &= - \frac{\langle (\partial_x u_x)^3 \rangle}{\langle (\partial_x u_x)^2 \rangle^{3/2}} \ ,
\end{split}
\end{equation}
where $S_0$ is the negative velocity derivative skewness, $Z(t)$ is the enstrophy 
and $P(t)$ is the palinstrophy.
For example, in 2D there is zero skewness and hence no vortex stretching.
From this equation, we
see that a larger skewness results in a greater amount of
vortex stretching, which may extend to higher dimensions.

In Figure \ref{fig:skewness} we show $S_0$ for 2D, 3D, and 4D.
The exact dependence of $S_0$ on Re$_{\lambda}$ is not known, however, the Kolmogorov 1962 theory \cite{K62} predicts $a$Re$_{\lambda}^b$. This is consistent with our data where we find $b^{\mathrm{3D}} = 0.08\pm 0.01$ and $b^{\mathrm{4D}} = 0.13\pm 0.01$. As such, we find that in four dimensions $S_0$ depends more strongly on Re$_{\lambda}$ than in three dimensions.
As is seen in the plot, 2D has roughly $S_0 = 0$, 
which is consistent with the absence of vortex stretching.
However, in 3D and 4D, $S_0$ increases with Re$_{\lambda}$, with skewness higher in 4D than 3D 
a trend that has been also been observed for free decay \cite{gwsns07}.
This data also suggests that if, as is predicted by the K41 theory, the skewness takes on a 
universal value as Re$_{\lambda}$ $\rightarrow \infty$ then this asymptotic value is larger in 4D.
We can then interpret this larger skewness value in 4D as being indicative of an enhanced rate 
of enstrophy production. Furthermore, this result can be understood as the extra spatial 
dimension allowing for additional vortex stretching, and hence an increased forward cascade of 
energy.

\begin{figure}
\includegraphics[width=0.5\textwidth]{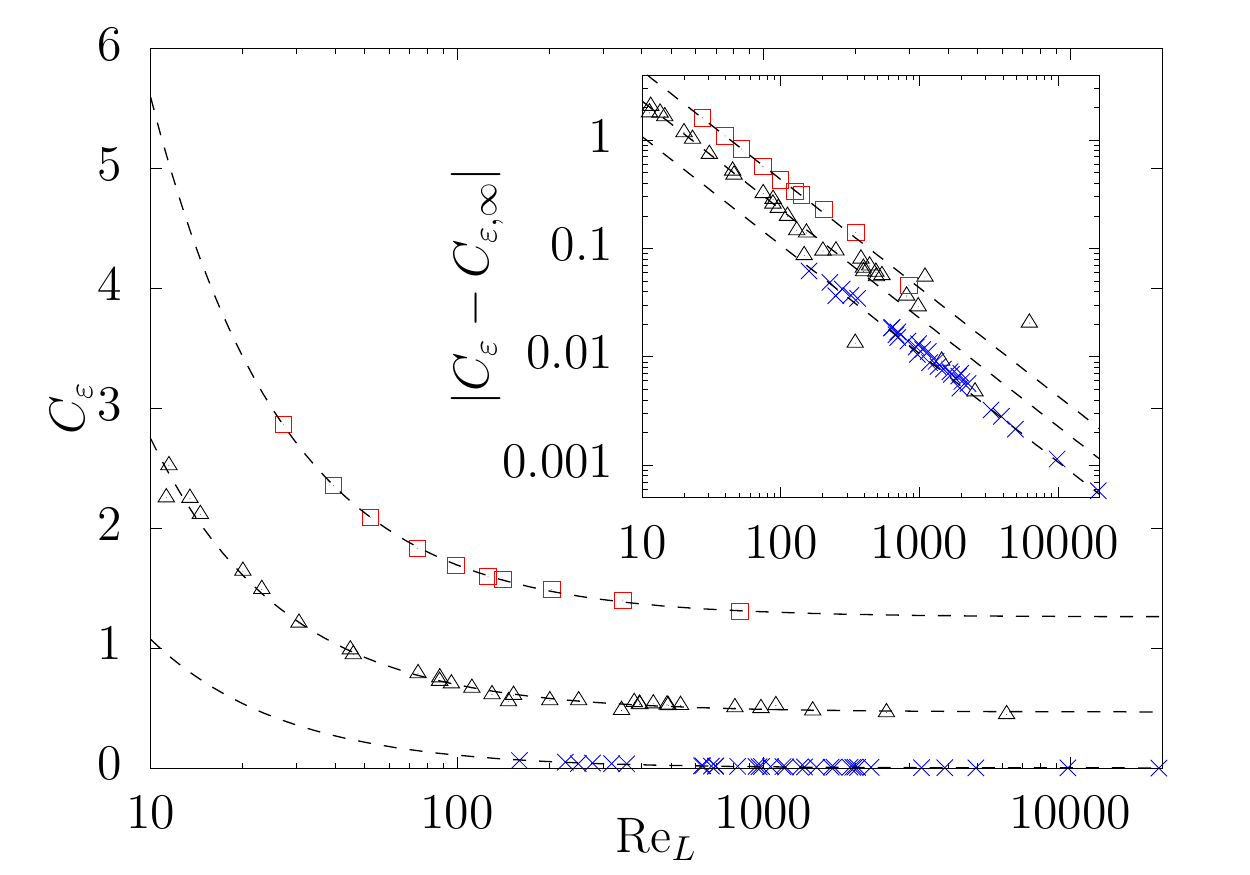}
\caption{\label{fig:Ceps} Main panel: $C_\varepsilon$ against Re$_{L}$.
Dashed line fit for model Equation (\ref{dimdisseq}) for two ($\times$), three ($\vartriangle$) and four ($\square$) dimensions.
Inset: $|C_\varepsilon - C_{\varepsilon,\infty}|$ against Re$_{L}$, dashed
line corresponds to power law behavior.}
\end{figure}

\begin{figure}
\includegraphics[width=0.5\textwidth]{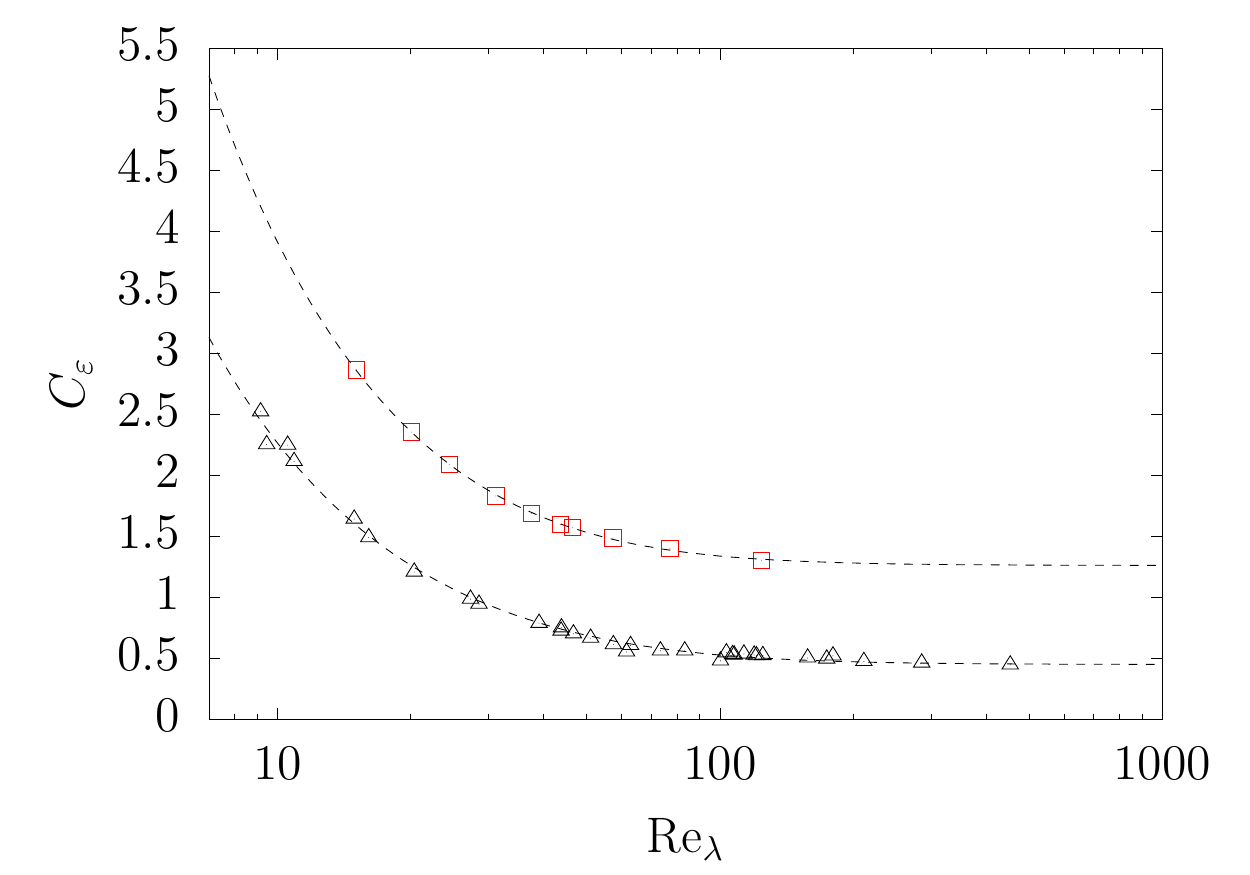}
\caption{\label{fig:Cepslambda}$C_\varepsilon(\mathrm{Re}_{\lambda})$ against Re$_{\lambda}$.
Dashed line fit for model Equation (\ref{dimdisslambda}) for three ($\vartriangle$) and four ($\square$) dimensions.}
\end{figure}

The dissipative anomaly in 3D turbulence, where the rate of energy 
dissipation tends to a nonzero asymptotic value in the 
limit of infinite Reynolds number, is one of the fundamental
phenomenological characteristics of turbulence. It clearly
distinguishes 3D from 2D dynamics and it is connected with
mathematical difficulties in proving regularity in the 3D
Navier-Stokes equations. The dimensionless dissipation rate is defined as
\begin{equation}
%\varepsilon = C_\varepsilon \frac{u^3}{L} \ .
	C_\varepsilon = \frac{\varepsilon L}{u^3} \ ,
\end{equation}
%is a well-studied quantity in 3D turbulence. 
There is ample experimental \cite{sreenivasan1, sreenivasan2, burattini}
and numerical \cite{wang1, gotoh1, donzis, bos1, yeung1, yeung2, ishihara} evidence
that, in 3D, $C_\varepsilon \to C_{\varepsilon,\infty} \neq 0$ 
as Re$_{L}$ $\ \rightarrow \infty$, indicating the persistence of a finite 
rate of energy dissipation even in 
the limit of zero viscosity. 
This is known as the dissipative 
anomaly and 
can be understood as being a consequence of vortex stretching 
\cite{Doering2009}
and thus of non-zero skewness. 
The dependence of $C_\varepsilon$ on Re$_{L}$ can be approximately described
as \cite{Doering2002,McComb2015}
\begin{equation}
\label{dimdisseq}
C_\varepsilon = C_{\varepsilon,\infty} + \frac{C}{\mathrm{Re}_{L}} \ ,
\end{equation}
where $C$ is a constant. A similar result can also be derived \cite{Doering2002,McComb2015} for $C_{\varepsilon}$ in terms of Re$_{\lambda}$, which gives \begin{equation}\label{dimdisslambda}
C_{\varepsilon}(\mathrm{Re}_{\lambda}) = A\left[1+ \sqrt{1 + \left(\frac{B}{\mathrm{Re}_{\lambda}}\right)^2} \right],
\end{equation} where $A$ and $B$ are constants with respect to Re$_{\lambda}$.

The value of $C_{\varepsilon,\infty}$ depends on dimensionality and 
the inviscid invariants, in 3D the high levels of helicity reduce 
$C_{\varepsilon,\infty}$ \cite{Linkmann18}, and in 2D, where there is no forwards cascade of energy, $C_{\varepsilon,\infty} = 0$. Hence it is of interest to 
examine the behavior of $C_\varepsilon$ also in 4D.
In Figure \ref{fig:Ceps} we show the Re dependence of $C_\varepsilon$ for 2D, 3D, and 4D
data, and we find that it is well described by Eq.~\eqref{dimdisseq} in all cases, 
albeit with different values of $C_{\varepsilon,\infty}$ and $C$.
Consistent with our results showing an enhanced forward cascade, we see an increase in the value of $C_{\varepsilon,\infty}$ with increasing dimension,
this grows from 0.467 for 3D in our data to 1.261 for 4D.
Since $C_{\varepsilon}$ is defined in terms of $u$ and $L$, which have explicit dimensional dependence, it may have been possible that the increasing value of $C_{\varepsilon,\infty}$ is solely due to changes in these quantities. 
However, our results for energy spectra and skewness are independent of how length and velocity scales are defined. Furthermore, the increase of $C_{\varepsilon,\infty}$ between 3D and 4D is greater than would be expected if it were solely caused by these dimensional dependences, thus we conclude that the increased asymptotic dissipation rate is a real effect. In Figure \ref{fig:Cepslambda} we show the dimensionless dissipation rate in terms of Re$_{\lambda}$ and find that the constants in equation \ref{dimdisslambda} take on the values $A =0.225342 $ and $B = 90.0105$ in 3D and $A = 0.63123 $ and $B = 51.0446$ in 4D.

Fluctuations are another important measure for assessing
change in behavior.
For the case of critical phenomena above the upper critical dimension, 
mean field theory becomes exact close to the critical point. 
If there is an upper critical dimension, then there should be true scale
invariance in the inertial range. This would result in reduced fluctuations and
smaller deviations from the K41 theory in terms of structure functions as
we move towards this dimension.
At criticality, the two point correlation functions of such systems 
display scale invariance.

\begin{figure}
    \includegraphics[width=\columnwidth]{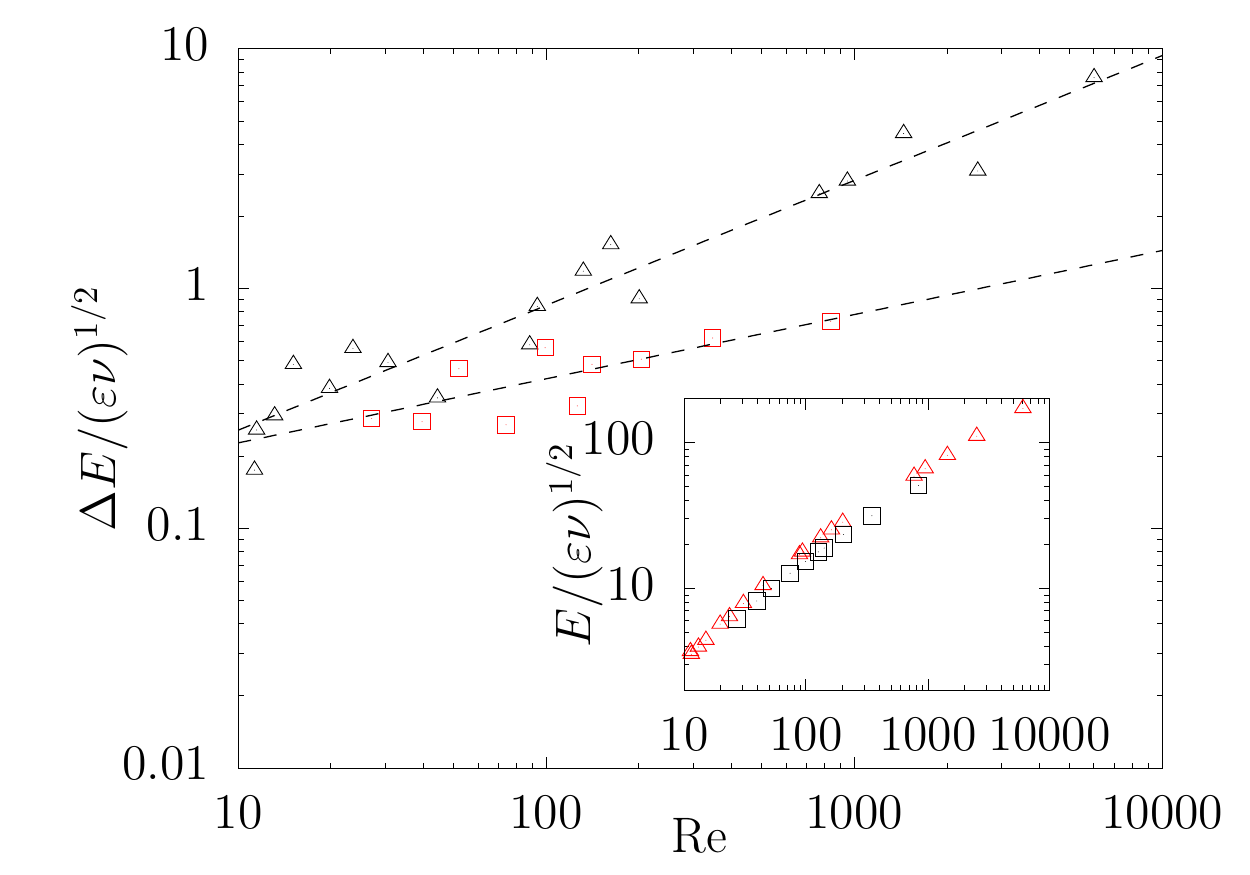}
    \caption{Normalized variation of energy with Re$_{L}$ in three ($\vartriangle$) and four ($\square$) dimensions. Inset: normalized energy with Re$_{L}$ at same scale as main plot.}
    \label{fig:energyvar}
\end{figure}

Figure \ref{fig:energyvar} shows a plot of $\Delta E/(\varepsilon \nu)^{1/2}$ with Re$_{L}$, 
where $\Delta E$ is the standard deviation of the total energy in time. 
This figure shows a measure of the fluctuation with Re$_{L}$ in 3D and 4D.
The dashed line shows a power law fit for illustrative purposes.
The fluctuations for 4D are smaller than 3D and rise slower.
The inset shows normalized energy with Re$_{L}$, with both 3D and 4D data being roughly similar. Thus, the decreased fluctuations are not merely an effect of
there being lower total energy.
One may also plot $\Delta E/E$ and the 4D case has values clearly lower than
the 3D values, with the difference becoming more pronounced at higher Re$_{L}$.
Another aspect of the simulations which suggests smaller fluctuations for 4D than in 3D, 
was the tendency of 4D simulations to reach statistically steady states 
in as little as half the time. Further to this, 
the fluctuation with wave number of the transfer spectra $T(k)$ were much greater
in 3D than 4D (not shown).

\section{Concluding remarks}
\label{concl}

This paper has reported a series of 4D HIT DNS simulations
for the forced Navier-Stokes equation.  
There have only been a few previous DNS simulations in 4D
\cite{Suzuki2005,gwsns07,ysing12,nikitin11}, 
all done for free decay.  These were groundbreaking
papers for this computationally demanding direction, but
free decay for the relatively small box sizes they achieved limited the 
period of fully-developed turbulence to be very short if at all.
Our simulations were run
for adequately long time and for sufficiently large box sizes
to produce for the first time a reliable and robust regime of
fully developed turbulence.  This now opens
the possibility to numerically test theoretical ideas about
4D turbulence
with a reliable 4D simulated turbulent state. 
As discussed  at the start of the paper,
there is various discussion scattered in the literature 
over many years
on how studying 4D turbulence might shed new insights in
the theoretical understanding of turbulence.
This work helps move one step further in that direction.

The numerical demands to simulate fully-developed turbulence in 4D
limits the extent of measurements that can be achieved.
Our results are modest but interesting for two main reasons.
Firstly, in 4D we find the presence of a seemingly enhanced forward 
energy cascade, consistent with what has been suggested in theoretical studies. Our results also show an increase in the asymptotic  dimensionless dissipation rate and velocity derivative skewness, which is further evidence of the enhanced forward energy cascade. Secondly, we see a reduction in the size of fluctuations in the total energy when going from three to four dimensions.
The reduction of these fluctuations 
is due to the non-linear transfer of energy between different 
length scales in the flow,
coming from an increased tendency of energy passing from
large to small scales. Noting that the reduction 
in Fig. \ref{fig:energyvar}
is on a log-plot, it is quite a dramatic decrease
in going from 3D to 4D (as is also the case for
$\Delta E/E$ which is not shown but we have checked).  Thus turbulence joins
critical phenomenon and QCD, as discussed in the Introduction,
as another strong coupling, multi-degree of freedom problem
that exhibits noteworthy changes from 3D to 4D.

For several decades the question has
lingered in the literature as to whether there are any
distinct differences to turbulence in three versus four
spatial dimensions.  
The barriers to answering this question have been to identify
appropriate quantities to measure and then measure them
to adequate computational reliability.  The handful of past
datasets \cite{Suzuki2005,gwsns07,ysing12} already
produced some interesting results that showed differences
between three and four dimensions. However these were small
datasets for which it is unclear
the degree to
which they realize fully developed turbulence.

In this work we have identified two
measures, one related to dissipation and another to fluctuations,
to compared between 3D and 4D turbulence. 
We have then developed a dataset at
adequately high resolution and evolved long enough, 
so as to realize fully developed
turbulence in 4D, from which we could then reliably take measurements
of these two quantities.  Our results have therefore provided
the first
definitive measurements of a 4D turbulence state,
from which we could demonstrate some clear
differences in turbulence between three and four spatial dimensions.
We do find some differences in the behavior of turbulence,
with, in particular, significant suppression of at least this one
measure of fluctuations in four compared to three dimensions.

These measurements are computationally very demanding,
as they are in four spatial dimensions,
need adequately high spatial and temporal resolution,
and require well equilibriated forced simulations.
Further definitive measurements of other forms of fluctuation
and dissipation  behavior in four and even higher dimensions
would be of interest.
The new insights learned from such efforts may assist
in reaching the long sought for theory of turbulence.
Nevertheless for now the computational demands place considerable
limitations on any rapid progress along these lines.

For instance, the measurements of the velocity-gradient skewness presented in Fig.~4 show
that extreme fluctuations in the velocity-field gradients become more likely in
4D than in 3D with increasing Reynolds number and thus at smaller and smaller
scale. This motivates fundamental questions concerning self-similarity that are
usually assessed in terms of structure function scaling, in particular at high
order, where deviations from dimensional scaling are observed in 3D. Such
measurements are very challenging in 4D, as they require an extended scaling
range in order to be reliable and this highly resolved simulations.  Our
results provide a first step in this direction and a motivation to take this
challenge on.

\begin{acknowledgments}
The Authors thank Moritz Linkmann for numerous helpful discussions and suggestions. This work has used resources from ARCHER \cite{archer} via the Director's Time budget.
This work used the Cirrus UK National Tier-2 HPC Service at EPCC \cite{cirrus} funded
by the University of Edinburgh and EPSRC (EP/P020267/1).
A.B. acknowledges funding from the 
U.K. Science and Technology Facilities Council,
R.D.J.G.H is supported by the U.K. Engineering and Physical Sciences Research Council (EP/M506515/1), and
D.C. is supported by the University of Edinburgh.
\end{acknowledgments}

\end{document}